\documentclass[usenatbib]{mnras}
\usepackage{newtxtext,newtxmath}

\usepackage[T1]{fontenc}

\DeclareRobustCommand{\VAN}[3]{#2}
\let\VANthebibliography\thebibliography
\def\thebibliography{\DeclareRobustCommand{\VAN}[3]{##3}\VANthebibliography}

\usepackage{graphicx}	
\usepackage{amsmath}	

\def\bi#1{\hbox{\boldmath{$#1$}}}

\newcommand{\mpc}{\ensuremath{\, h^{-1}\,\mathrm{Mpc}\, }}
\newcommand{\mpccube}{\ensuremath{\, h^{-3}\,\mathrm{Mpc}^3 }}

\newcommand{\lya}{Ly$\alpha$}

\newcommand{\beq}{\begin{equation}}
\newcommand{\eeq}{\end{equation}}
\newcommand{\bc}{\begin{center}}
\newcommand{\ec}{\end{center}}
\newcommand{\bfig}{\begin{figure}}
\newcommand{\efig}{\end{figure}}

\title{Joint Cosmic Density Reconstruction from Photometric and Spectroscopic Samples}

\author[B. Horowitz et al.]{
B. Horowitz,$^{1}$\thanks{E-mail: bhorowitz@princeton.edu}
P. Melchior$^{2,3}$
\\
$^{1}$Lawrence Berkeley National Lab, 1 Cyclotron Road, Berkeley, CA 94720, USA\\
$^{2}$Department of Astrophysical Sciences, Princeton University, Princeton, NJ 08544, USA\\
$^{3}$Center for Statistics \& Machine Learning, Princeton University, Princeton, NJ 08544, USA
}

\date{Accepted XXX. Received YYY; in original form ZZZ}

\pubyear{2023}

\begin{document}
\label{firstpage}
\pagerange{\pageref{firstpage}--\pageref{lastpage}}
\maketitle

\begin{abstract}

We reconstruct the dark matter density field from spatially overlapping spectroscopic and photometric redshift catalogs through a forward modelling approach. Instead of directly inferring the underlying density field, we find the best fitting initial Gaussian fluctuations that will evolve into the observed cosmic volume. To account for the substantial uncertainty of photometric redshifts we employ a differentiable continuous Poisson process. In the context of the upcoming Prime Focus Spectrograph (PFS), we find improvements in cosmic structure classification equivalent to 50-100\% more spectroscopic targets by combining relatively sparse spectroscopic with dense photometric samples.

\end{abstract}




\section{Introduction}

For the past half century, spectroscopic galaxy redshift surveys have been the key method used to study the three-dimensional distribution of matter in our universe \citep{gregory1978coma,kirshner1978study,davis1982survey}. This intricate structure, known as the cosmic web, consists of under-dense void regions enclosed by sheet-like walls, which are embedded between filamentary structures that connect dense dark matter halos of galaxy clusters \citep{1996Natur.380..603B}. Signatures of these structures are well established (e.g. \cite{2011MNRAS.416.2494P,2014MNRAS.438.3465T}), and there is significant interest in connecting galaxy properties, like inferred star formation rate, with their cosmic environments \citep{2017A&A...597A..86P,2022arXiv220614908G}.

Photometric surveys, on the other hand, provide significant cosmological information through their clustering statistics \citep{2020JCAP...03..044N}, gravitational lensing effect \citep{2013MNRAS.430.2200K}, and cross correlations with spectroscopic catalogs \citep{2015MNRAS.449.1352C}.
Upcoming surveys, such as the Legacy Survey of Space and Time \citep[LSST,][]{2009LSST}, and ongoing surveys like the Hyper Suprime-Camera (HSC) on the Subaru Telescope are greatly expanding the number of galaxies with multiwavelength data suitable for photometric analysis \citep{2021MNRAS.500.1003W}. 
However, the substantial uncertainty in their inferred redshifts (``photo-z'') is a major limitation for their utility in tracing cosmic structure.



The past few years has seen an explosion of work in foreword modelling of galaxy density fields and applying those techniques to real survey data. These include the Bayesian Origin Reconstruction from Galaxies method \citep[BORG]{2013BORGI} applied to the SDSS/BOSS galaxy survey \citep{2019BORGII}, and the more recent COSMological Initial Conditions from Bayesian Inference Reconstructions with THeoretical models \citep[COSMIC BIRTH]{2020BIRTHI} applied to the COSMOS field \citep{2021BIRTHII}. Both these techniques rely on computationally expensive Hamiltonian Monte Carlo (HMC) sampling techniques, requiring on the order of 10,000 samples, each requiring a number of forward model computations. An alternative method was proposed in \cite{seljak2017towards}, where the authors pointed out that a Gaussian approximation for the likelihood is fairly accurate for cosmological analysis, and a simple quasi-Newtonian optimization scheme can achieve accurate results in $\mathcal{O}(100)$ steps. An implementation of this method was first applied to a realistic galaxy mock catalog in \citet{2021TARDISII}, finding an accurate reconstruction of the cosmic web. These techniques have since been combined with Lyman-$\alpha$ forest measurements and been used for map making from data from the COSMOS Lyman Alpha Mapping And Tomographic Observations\citep{2021arXiv210909660H}. 

A critical aspect for both the quasi-Newtonian optimization and the HMC analysis is the model to map from the evolved dark matter distribution to the observed galaxies which act as biased tracers. These methods must be fully differentiable; i.e. one must be able to calculate the derivative of the final (gridded) tracer density field with respect to the initial matter density field and/or particle positions. While their have been a number of fast particle-mesh solvers developed over the past decade \citep{2005NewA...10..393M,2013JCAP...06..036T,2016MNRAS.459.2327I}, only a few have been designed with integrated derivative calculations \citep{fastPM,2020arXiv201011847M}. In addition, one must still further map from the evolved density field (or the found displacement field) to the tracer density in a differentiable way.

In this work, we present an approach to jointly modelling both spectroscopic and photometric redshift catalogs with the goal of cosmic web reconstruction. We extend the forward galaxy modelling framework in \citet{2021TARDISII} to account for significant uncertainty in galaxy positions through a  Poissonian likelihood. We present our likelihood model and mock catalog construction in Section \ref{sec:methods}. We show our reconstructions of the cosmic web and compare with our simulated truth in Section \ref{sec:results} and finally discuss the results and further applications in Section \ref{sec:conclusion}. Throughout this work we assume a Planck 2015 cosmology as discussed in \cite{2016A&A...594A..13P}.

\section{Methods}
\label{sec:methods}

\subsection{Dynamical Forward Model}

We follow past works \citep{2019TARDIS,2021TARDISII} and use a dynamical model to forward simulate our observed universe and iteratively find the maximum a posteriori initial conditions. 
In short, we solve for initial density field, $\bi{s}\in\mathbb{R}^M$, given some observed data $\bi{d}\in\mathbb{R}^N$. We can map our initial density field to the data with a forward operator, $\bi{f}$.
In general this forward operator will depend on additional parameters (i.e. cosmological or astrophysical parameters) which will need to be marginalized over depending on the final quantity of interest. For this work we assume these are fixed (see \citet{seljak2017towards} for a discussion of this marginalization). 

This is a highly dimensional optimization problem, approximately equal to the number of initial particles in our reconstruction simulation. To use gradient based optimization methods, like L-BFGS, we need to have a fully differentiable forward model to map from our initial density field to the late time observed quantities. The dynamical evolution of the field itself is done with FlowPM \citep{2020arXiv201011847M}, a tensorflow implementation of the FastPM \citep{fastPM} package. This implementations has the advantage of GPU-compatibility while also being completely differentiable. The FastPM framework has been well tested to reproduce halo statistics, power spectrum, and correlation functions as more traditional n-body solvers (TreePM). We use 8 timesteps to evolve the field from $z=10$ to $z=1.0$, linearly spaced in scale factor, with a force resolution of 1. We choose $z=1.0$ as it is the mean target redshift for the PFS-Galaxy Evolution low-$z$ survey, the main motivation for this work.

For our differentiable galaxy forward model we use second order Lagrangian bias using the displacement, $\bi{\psi}$, calculated from our particle-mesh simulation. In this framework, the galaxy field $\delta_g$ can be expressed as a
\begin{equation}
    1+\delta_g(\bi{x}) = \int d^3 \bi{q} w(\bi{q})\delta_D(\bi{x}-\bi{q}-\psi),
\end{equation}
where $w(\bi{q})$ is a weight function (i.e. response) expressed as a sum of of the Lagrangian density as 
\begin{equation}
    w(\bi{q}) = 1 + b_1\delta(\bi{q}) + b_2 \left[\delta^2(\bi{q})-\langle \delta^2(\bi{q}) \rangle \right].
\end{equation}
Note that, in practice, this integral expression can be done trivially at the field level via painting our weighted field on the initial (i.e. Lagrangian) density field and applying displacement calculated by the dynamical forward model. Additional bias terms (such as tidal shear) could be added if they can be expressed as differentiable operators of outputs from our particle mesh simulation. We calibrate $b_1$ and $b_2$ by fitting the mock observed power spectra to that from a separate mock catalog (see \cite{2021TARDISII} for additional discussion of this approach).  Bias parameters could also be fitted jointly with the optimization, with an increased computational cost and possible decrease in reconstruction accuracy. 


\subsection{Galaxy and Photometric Redshift Likelihood Model}
\label{subsec:gl}

At each step in our optimization, we need to compare our reconstructed dark matter field with the (true) observed photometric galaxy catalog. In \citet{2021TARDISII}, the authors performed an density reconstruction of spectroscopic observed galaxies. In that work the authors used a $\ell_2$ norm for comparing the observed galaxies with a forward modelled bias density field. As studied in \cite{2021JCAP...03..058N}, a Poisson likelihood term provides a more accurate reconstruction of small-scale power and, in the case of coarse galaxy resolution (i.e. $\lesssim 5$ galaxies per grid square) the Gaussian approximation is arguably no longer mathematically valid.

With this in mind, we model our observed catalog as a draw from the inhomogeneous Poisson process in $\bi{x}\in\mathbb{R}^3$, defined by the observable rate density $\Phi(\bi{x}\mid \bi{s}) = f(\bi{x}\mid \bi{s})\mathcal{W}(\bi{x})$, a product of the sample density predicted by the forward model and the survey detection efficiency $\mathcal{W}$.

We can write the probability of the observed data for $K$ objects, $\bi{d} = \{\bi{x}_1, \cdots, \bi{x}_K\}$, given some initial density field as
\begin{equation}
    \mathcal{P}(\bi{d} \mid \bi{s}) = \exp\left(-\int d^3x\, \Phi(\bi{x}\mid\bi{s})\right)\prod_{k=1}^{K}\Phi(\mathbf{x}_k\mid\bi{s}).
    \label{eq:prob}
\end{equation}
In this formula, the exponential term acts as a normalization to express the expected total number of observations. As our forward model is binned in cells, we transform the product over galaxies to a product over grid cells:
\begin{align}
\log\mathcal{P}(\bi{d} \mid \bi{s}) &= \sum_i n_i \log \phi_i - \phi_i\Delta_i^3\mathrm{, with}\\
n_i & =\sum_k \iota(x_k \in \mathcal{C}_i)\ \mathrm{and} \label{eq:ni}\\
\phi_i &= \int_{\mathcal{C}_i} d^3x\, \Phi(\bi{x}\mid \bi{s})
\end{align}
being the number of observed samples that fall in cell $\mathcal{C}_i$ and the total expected rate in that cell, respectively. $\Delta^3_i$ is the cell volume.

\autoref{eq:ni} requires that we can directly count the number of times a galaxy falls into a specific cell, which implies positional uncertainties much smaller than the size of the cells. This requirement is concerning already for spectroscopic samples, but it cannot be met with photometric ones with dozens of Mpc uncertainties in the redshift direction.
We can generalize the number counts to account for the redshift uncertainty $p(z_k)$ of each galaxy in that cell:
\begin{equation}
\label{eq:ni_prime}
    n^\prime_i = \sum_k \mathcal{P}\left(x_k\in\mathcal{C}_i\mid p(z_k)\right).
\end{equation}
This amounts to a redistribution of the counts according to the probability mass every $p(z_k)$ assigns to each cell.
If all galaxies are assumed to have the same redshift uncertainty, \autoref{eq:ni_prime} simplifies to a one-dimensional convolution of the observed counts with the redshift kernel in the radial direction, but our approach does not require this assumption.
Because the modified counts are no longer integer, we employ the generalization of the Poisson distribution for continuous variables, which can be expressed in terms of the incomplete Gamma function probability:
\begin{equation}
\mathcal{P}(n^\prime_i \mid \bi{s}) = \frac{1}{\Gamma(n^\prime_i)} \int_0^{\phi(\bi{s})_i} t^{n^\prime_i-1}e^{-t} dt,
\end{equation}
where $\Gamma$ is the normal gamma function. 

\begin{figure*}
    \centering
    \vspace{-1.5cm}
    \includegraphics[width=\textwidth]{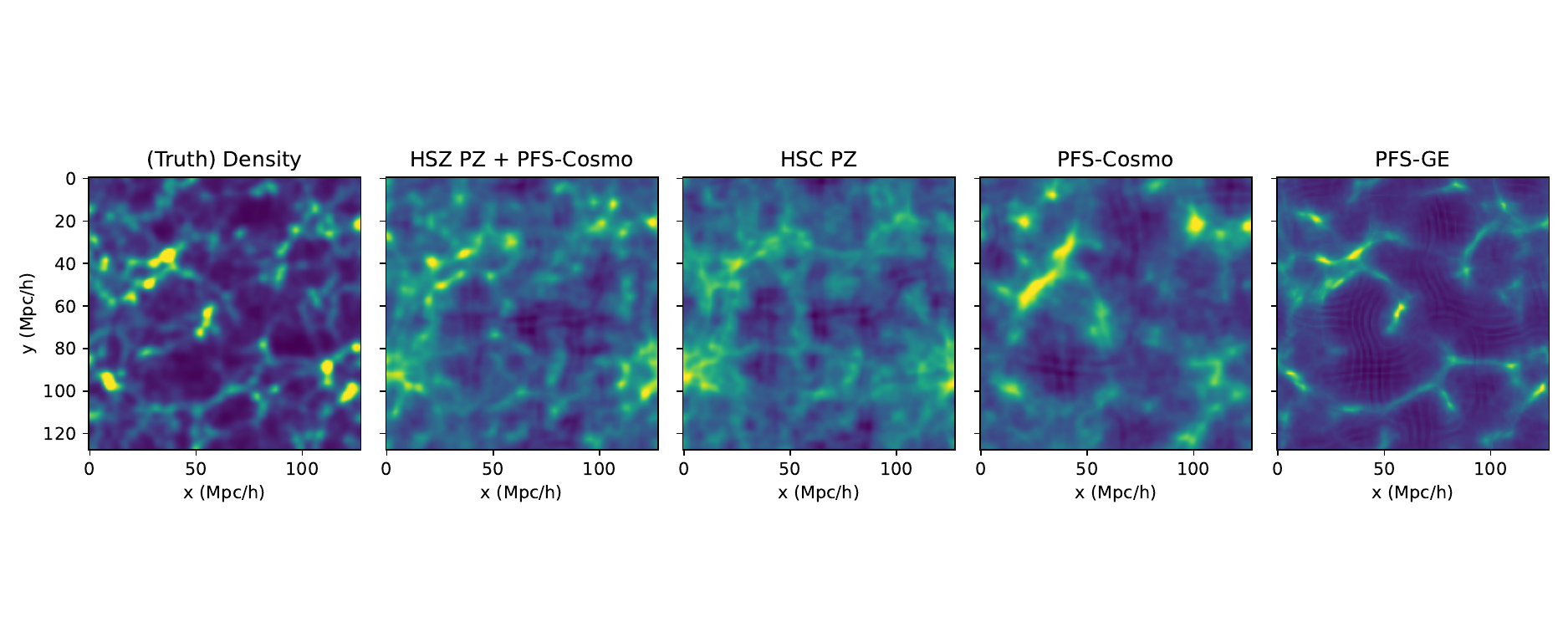}
    \vspace{-1.5cm}
    \caption{Projection along the $z$ axis for a 10 $h^{-1}$ Mpc slice of a simulated volume as reconstructed by a mock spectroscopic (PFS-Cosmo), photometric (HSC), and jointly both surveys. We find that the photometric reconstruction provides an accurate but diffuse map of the underlying structures, while the spectroscopic reconstruction is able to pick out the most massive clusters but lacks the number density to follow the low-density structures. The joint reconstruction is able to capture both features and shows good qualitative agreement to the true density field.}
    \label{fig:3-panel}
\end{figure*}

\begin{table*}
\centering
\begin{tabular}{llll}
               & HSC & PFS-Cosmo & PFS-GE \\
               \hline
Number Density (h$^{-1}$ Mpc)$^{-3}$  & 0.06   & 0.0006 & 0.025                    \\
Redshift Error (h$^{-1}$Mpc) & 30.0   &  0.5           & 0.5                   
\end{tabular}
\centering
\caption{Configurations chosen to mimic the upcoming Prime Focus Spectrograph (PFS) survey and the Hyper Suprime-Cam (HSC) photometric survey.}\label{tab:config}
\end{table*}

\subsection{Mock Catalog and Redshift Error Modeling}

We construct a mock simulated volume using FastPM, which has been well tested to reconstruct proper halo statistics and profiles for the range of interest in this work \citep{fastPM}. We use a simulated volume of $128^3$ \mpccube with particle resolution of $1024^3$, with time resolution of 10 steps. We use a friends of friends halo finder from Nbodykit \citep{hand2018nbodykit}, finding 2,097,149 halos in our volume. We use the Zheng et al. \citep{2005ApJ...633..791Z} HOD model from halotools \citep{2016MNRAS.460.2552H} with associated HOD parameters $\log{M_\textrm{min}}=12.02$, $\sigma_{\log{M}}=0.26$, $M_0 = 11.38$, $\log{M_1}=13.31$ and $\alpha = 1.06$, matching those found in \citet{2007zheng}. This results in 91588 central and satellite galaxies, from which we down-sample to form our mock catalogs for each survey.

For our mock catalog we model the Prime Focus Spectrograph's cosmology survey or PFS Galaxy Evolution survey combined with the overlapping HSC targetting photometric catalog. The PFS Cosmology survey is designed to reconstruct the growth rate of structure and geometry via a variety of probes on large scales (Bayronic acoustic oscillations, redshift space distortions, clustering) \citep{takada:2014} while the Galaxy Evolution survey is currently designed to reach a significantly larger number density in order to resolve cosmic web structures and identify galaxy groups and clusters \citep{greene2022prime}. At $z \sim 1$, each tasks require different number densities of spectroscopic targets, and we show the corresponding number densities for each sample in Table \ref{tab:config}.

We model our photometric redshift errors inspired by the average photo-z calibration error described in HSC PDR2 \citep{2020arXiv200301511N}. We assume that all galaxies have a similar photometric error distribution, with Gaussian error of 30 Mpc/h corresponding to a $\Delta z \sim 0.015$ at $z=1.0$. More complex photo-z distributions can be included in a straightforward fashion via re-expression as a Gaussian mixture model and adjusting the rate density in Eq. \ref{eq:prob2}. When combining HSC with a spectroscopic data set, we assume all spectroscopically identified galaxies are also in the HSC catalog and remove them in order to not double count a given galaxy. 

We assume all galaxies have the same bias properties, but it would be simple within our formalism to use multiple different types of biased galaxy subsets, each with their own log likelihood that is then added together. We smooth the observed three-dimensional galaxy field by 0.05 $h^{-1}$ Mpc because sharp discontinuities can lead to a ``stuck" optimization at a non-optimal solution.

\subsection{Optimization Method and Prior}

Our method relies on an high-dimensional optimization for the underlying initial density field, $\bi{s}\in\mathbb{R}^M$. The probability of some given observed data $\bi{d}$ can be expressed as
\begin{equation}
    \mathcal{P}(\bi{d}) = \int d\bi{s} \mathcal{P}(\bi{s}) \mathcal{P}(\bi{d}\mid\bi{s}),
\end{equation}
where $\mathcal{P}(\bi{s})$ is a prior on the initial density field and $\mathcal{P}(\bi{d}\mid\bi{s})$ is the likelihood with respect to the fields. This expression can further be constrained by imposing a constraint on the signal field based on cosmological model. Assuming the initial density is a Gaussian random field described by covariance matrix, $\bi{S}$, we can write down the prior as 
\begin{equation}
    \mathcal{P}(\bi{s}\mid\bi{S}) = \frac{\exp{\left(-\frac{1}{2}\bi{s}^\dag \bi{S}^{-1}\bi{s}\right)}}{\det{2\pi \bi{S}}},
\end{equation}


Usually the dimension of the initial field, $M$ will be quite large, resulting in a very high-dimensional inverse problem.  For example, in \citet{2019TARDIS} this procedure was used to reconstruct the initial density field from Lyman Alpha forest measurements and required $M = 128^3$ for an accurate reconstruction over a $64$ $h^{-1}$ Mpc box. Optimization over such a large number of parameters would be impractical without using derivative information to inform each update in the optimization. We use an Limited Memory BFGS optimization method, implemented in scipy, which uses derivative information at each step to update a Hessian approximation at each step (a ``quasi-Newtonian" method). 

In general there is no guarantee the likelihood surface is well behaved and convex. However, for the examples shown in this work we find little quantitative difference in our optimization by starting our optimization at different initial density fields. Alternative methods, like Hamiltonian Monte Carlo, could be used if this is a major concern \citep{2020BIRTHI,2013BORGI}. This is particularly true if one wants to use the reconstructed density field to update the priors of a redshift of a given galaxy (similar to clustering redshift calculations).



\section{Results}
\label{sec:results}
We have applied the reconstruction algorithm to our mock catalogs representing HSC and PFS-Cosmo and show a projected slice of the reconstructions at $z=1.0$ in Figure \ref{fig:3-panel}. Qualitatively, we find strong agreement between the fields, with structures in the reconstructions appearing more diffuse than the corresponding true structures. The spectroscopic reconstructions suffers from sparse sampling and can thus not recover smaller structures. In contrast, the higher-density photometric catalog allows to probe smaller and lower-density structures but lacks the spatial sharpness to accurately recover compact massive structures resulting in washed out features. 

\begin{figure}
    \centering
    \includegraphics[width=0.48\textwidth]{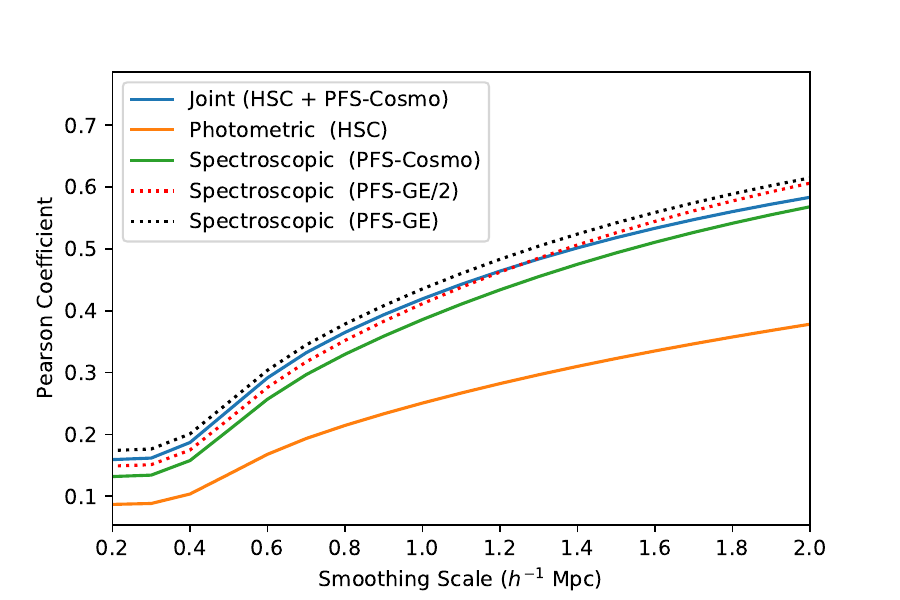}
    \caption{Pearson correlation coefficient as a function of smoothing scales between the reconstructed volume and the simulated truth. As we broaden the Gaussian smoothing kernel, we are able to match the simulated truth better, at the cost of small scale detail. We include reconstructions using half of the PFS-GE sample density and the full PFS-GE sample. See Table \ref{tab:config}. 
    }
    \label{fig:pc}
\end{figure}

For a quantitative comparison between the reconstructions at the field level we can compare the fields pixel by pixel via their Pearson correlation coefficient, defined as 
\begin{equation}
    r_{xy} =\frac{\sum ^n _{i=1}(x_i - \bar{x})(y_i - \bar{y})}{\sqrt{\sum ^n _{i=1}(x_i - \bar{x})^2} \sqrt{\sum ^n _{i=1}(y_i - \bar{y})^2}}.
\end{equation}
We plot this coefficient as a function of smoothing scale in Figure \ref{fig:pc} for both spectroscopic and photometric samples, as well as the joint analysis. We also show the result for using the higher density PFS-Galaxy Evolution survey as well as using half the density of that survey. We find, at a scale of 1 \mpc, the joint reconstruction performs comparably to the half of the PFS-Galaxy Evolution survey alone despite requiring significantly fewer spectroscopic targets (1260 vs. 26140).

\begin{figure*}
    \centering
    \vspace{-0.5cm}
    \includegraphics[width=\textwidth]{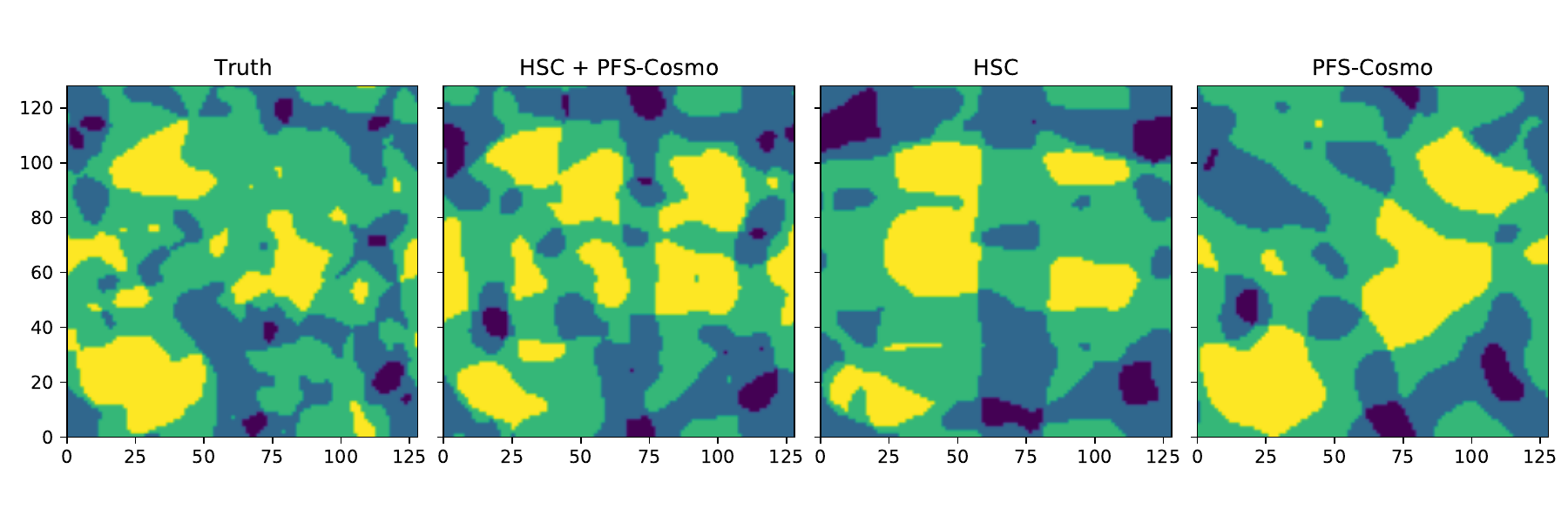}
    \vspace{-1cm}
    \caption{Qualitative cosmic structure plot for a sample x-z slice in the reconstructed volume classified using the deformation tensor approach for the spectroscopic redshift (PFS-Cosmo), photometric redshift (HSC) and joint surveys compared to the simulated truth. Dark blue corresponds to nodes (i.e. clusters), light blue to filaments, green to sheets, and yellow to voids.
    }
    \label{fig:cs_3p}
\end{figure*}

\begin{figure*}
    \centering
    \includegraphics[width=0.90\textwidth]{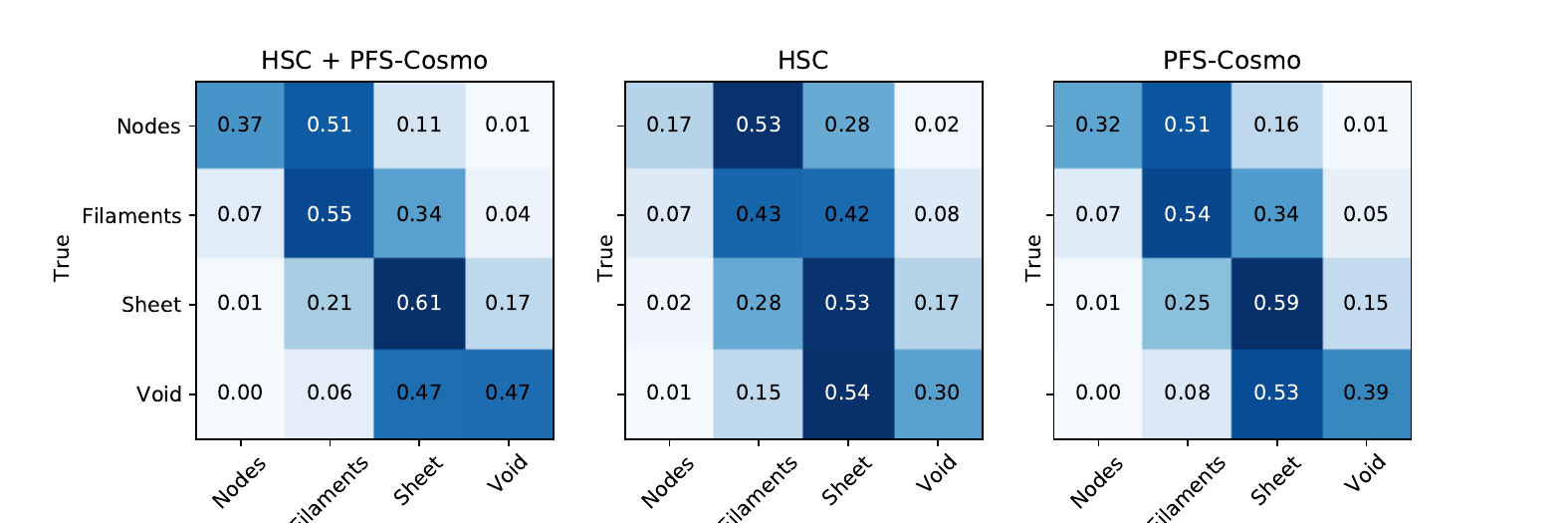}
    \caption{Cosmic web classification confusion matrix for our reconstruction for the spectroscopic redshift (PFS-Cosmo), photometric redshift (HSC) and joint surveys Perfect reconstruction would correspond to no off-diagonal terms. The joint reconstruction uniformly improves all cosmic structure classifications. Note that there is significant confusion between notes and filaments (and a lesser extent between sheets and voids) due to difficulty in exactly locating structures along the line of sight.}
    \label{fig:cs_cm}
\end{figure*}

\begin{figure*}
    \centering
    \includegraphics[width=0.90\textwidth]{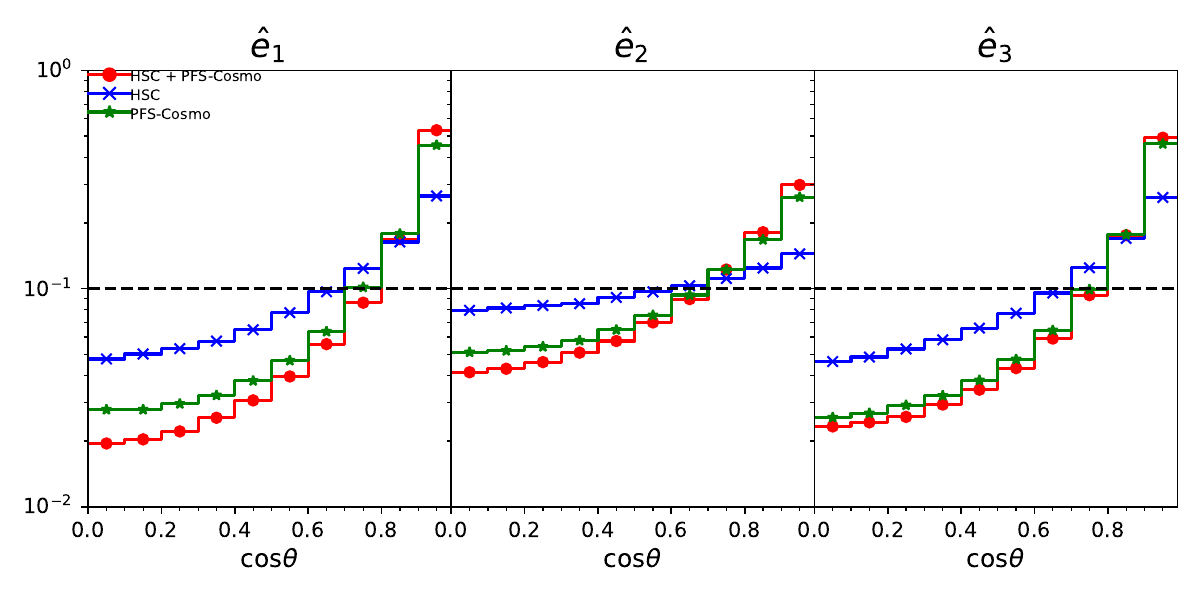}
    \caption{Histogram showing the eigenvector orientation recovery, corresponding to the principle inflow (outflow) directions of the cosmic structures, by computing the dot product of the reconstructed eigenvector with the true eigenvector. Perfect alignment corresponds to all weight in the far right side bin, and random alignment corresponds to the black dotted line. We also show (in red) the structured inferred by taking the spectroscopic galaxy field and smoothing it at the same resolution as the other fields, 2 $h^{-1}$Mpc.}
    \label{fig:align}
\end{figure*}

\subsection{Structure Classification: T-Web}
\label{subsec:structure}

Following \citet{2019TARDIS,2021TARDISII}, we use an the eigenvalues and vectors of the pseudo-deformation tensor, sometimes called the ``T-web" classification, as described in \citet{2016LeeWhite,2017Krolewski} and based on work in \citet{Bond:1993,2007Hahn,forero-romero:2009}. The pseudo-deformation tensor is the Hessian of the gravitation potential and its eigenvectors describe the principle inflow (outflow) directions assuming the corresponding eigenvalue is positive (negative) assuming the Zel'dovich approximation \citep{1970Zeldovich}. This can be efficiently computed in Fourier space with matter density, $\delta_k$, as
\begin{equation}
    \tilde{D}_{ij} = \frac{k_i k_j}{k^2}\delta_k.
    \label{eq:diften_k}
\end{equation}
We use this definition, as opposed to using the reconstructed nonlinear velocity field, in order to make a direct comparison to existing works in the literature which use this definition.

To classify cosmic structure into discrete classes we compare the eigenvalues to some thresh-hold value, $\lambda_{th}$. While one could use only the signature of the eigenvalues (i.e. $\lambda_{th} = 0$), this would result in a very low node fraction as only the pixel at local maxima would be considered true nodes. Following the approach in \cite{2019TARDIS,2021TARDISII}, we instead use set our thresh-hold for each map such that all our reconstructed fields have the same void volume fraction of 25\%. We use voids, rather than nodes, as they are less sensitive to the effective smoothing scale of the tracers. We show the classified fields in Figure \ref{fig:cs_3p}. We can compare pixel by pixel to compute a confusion matrix to see how often given cosmic structures are mis-classified, which is shown in Figure \ref{fig:cs_cm}. We find the joint reconstruction performs significantly better than the photometric catalog alone and moderately better than the spectroscopic catalog. In particular we find substantially fewer significant misclassifications (i.e. node misidentified as void or visa versa). 

In addition to discrete classification based on eigenvalues, the eigenvectors of the deformation tensor can be used to categorize the orientations of the cosmic structures. Each eigenvector indicates the principle inflow or outflow direction from the structure based on their signature. We can then determine the angle $\theta$ between these eigenvectors and the simulated true eigenvector, with the ideal result being a distribution that peaks at $\cos(\theta)=1$. We show this alignment in Figure \ref{fig:align}. We find the HSC photometric catalogs contribute moderately to the reconstructed alignment in the joint catalog with PFS-Cosmo but are poor tracers alone of the alignments.

\subsection{Halo Recovery}

\begin{figure}
    \centering
    \includegraphics[width=0.40\textwidth]{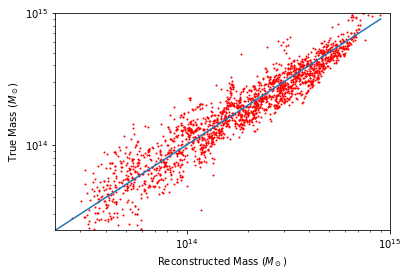}
    \caption{Reconstructed halo mass from the joint PFS-Cosmo + HSC catalog defined by a spherical overdensity vs. the true mass in the corresponding overdensity in our simulated density field. We find reasonable reconstruction of halo mass, with $\frac{\delta M}{M} =0.42$ for halos with $M> 2 \times 10^{13} M_\odot$.}
    \label{fig:Mass_Reconstruction}
\end{figure}

Instead of the continuous eigenvalue description used in Sec \ref{subsec:structure}, we can perform a discrete classification based on a halo-finding algorithm. This allows us to identify and easily compare individual cosmic structures, for example to identify galaxies to the same massive halo. 

There are a number of possible ways of determining halos, including using a friends-of-friends algorithm on the particle data (as done in \cite{2021TARDISII}) or using spherical over-density (for example in \cite{2015StarkProtocluster}).  
For this work, we use halos identified by a friends-of-friends algorithm (from nbodykit \citep{hand2018nbodykit}) from our reconstructed particle catalog. As we are interested in accuracy of the overall environment of galaxy clusters as opposed to construction of a specific halo catalog for cosmological analysis, we perform our comparison not on the friends-of-friends mass but on a spherical volume surrounding the center of each identified halo. This allows direct comparison of the halos identified in the reconstruction volume to those in the simulated truth without needing to account for specific halo identification between the two catalogs. This also minimizes the effects of minor variations in halo detection parameters causing a true halo to ``fragment'' into two smaller halos. We compare the recovered mass of our catalog to the true mass in Figure \ref{fig:Mass_Reconstruction}. We find $\frac{\delta M}{M} =0.42$ for halos with $M> 2 \times 10^{13} M_\odot$.

\section{Conclusion and Discussion}
\label{sec:conclusion}
In this work, we have presented the first joint analysis of a spectroscopic and photometric galaxy samples for reconstruction of the initial density field of an observed cosmic volume. We have shown that even a fairly sparse spectroscopic galaxy sample, when combined with a deep photometric galaxy catalog, can provide an accurate reconstruction of the cosmic web and recover major over-densities. We have demonstrated the competitive performance with the cosmic web classification by a deformation tensor approach as well as via the statistics of spherical over/under-densities. We have shown this method with emphasis on the upcoming Prime Focus Spectroscopic (PFS) Survey, combined with the Subaru Hyper Suprime-Cam surveys photometry. While cosmic web reconstruction is a the focus for the PFS - Galaxy Evolution survey, the method presented here is entirely general and could also be applied to combined the upcoming Legacy Survey of Space and Time (LSST) \citep{2009LSST} survey combined with a variety of spectroscopic surveys.

In \cite{2021TARDISII}, a similar approach was used to perform a joint reconstruction using \lya\ forest data and spectroscopic redshift catalogs. In that case the data was highly complementary and the reconstruction synergistic as each probe provided information on different spatial scales. In this work, we find the effects are more additive since the probes have comparable bias properties and probe comparable physical scales. In \cite{2021JCAP...10..056M}, the authors combined photometric surveys and intensity mapping in a comparable forward modelling framework. However, the focus of that work was on reconstructing large scales for cosmological analysis and didn't consider late-time cosmic web statistics, and used significantly different likelihoods due to that application. 

Recently, several works employed modern machine learning techniques to perform a more accurate photometric redshift prediction (e.g. \cite{2020MNRAS.497.4565E,2020RAA....20...89M,2021arXiv210104293A})\footnote{For an updated more complete list, see \url{https://github.com/georgestein/ml-in-cosmology}.}. Approaches that use individual galaxies photometry to infer redshifts can be trivially incorporated into these reconstruction techniques. More nuanced approaches that rely on correlations between galaxies, such as clustering redshift approaches, require more care as they likely will include correlations in the redshift errors which would need to be propagated through the likelihood function.

In this work we use a biasing scheme to map from density to galaxy field. While these approaches have been demonstrated to work well at describing structure on large scales, on small scales they are known to break down \citep{desjacques2018large}. An alternative approach could be to include in our forward model an explicit halo finding and halo population step. Differentiable versions of halo finding has been explored using neural network approaches in \citep{2018JCAP...10..028M,2020arXiv201011847M,2019PhRvD.100d3515K} and a differential HOD model has been developed in \citet{2022DHOD}. We plan to further explore applying these methods in upcoming work.

While not directly addressed in this work, the use of photometric redshift catalogs could help mitigate the effects of fiber collisions in cosmic web reconstructions \citep{2013AJ....145...10D,2014A&A...566A..84M}. In spectroscopic surveys covering dense cluster environments there is difficulty in achieving uniform sampling due to mechanical limitations of the fiber positioners. Since these environments are still well sampled by photometric surveys (assuming proper deblending), our reconstruction technique could substantially reduce the bias from this localized incompleteness.

\section*{Acknowledgements}

This work is supported by the AI Accelerator program of the Schmidt Futures Foundation. 

This research used resources of the National Energy Research Scientific Computing Center, a DOE Office of Science User Facility supported by the Office of Science of the U.S. Department of Energy under Contract No. DEC02-05CH11231.

This work made extensive use of scipy \citep{2020SciPy-NMeth} and Tensorflow \citep{tensorflow2015-whitepaper}.




\bibliographystyle{mnras}
\bibliography{example} 








\bsp	
\label{lastpage}
\end{document}